\documentclass[conference]{IEEEtran}

\usepackage{cite}
\usepackage{amsmath,amssymb,amsfonts}
\usepackage{algorithmic}
\usepackage{graphicx}
\usepackage{textcomp}
\usepackage{xcolor}
\def\BibTeX{{\rm B\kern-.05em{\sc i\kern-.025em b}\kern-.08em
  T\kern-.1667em\lower.7ex\hbox{E}\kern-.125emX}}
\usepackage{diagbox}
\usepackage{subcaption}

\begin{document}

\title{Student Competency Assessment and Presentation Methods Based on Algorithm Courses}

\iftrue

\author{
\IEEEauthorblockN{Yingqi Zhang}
\IEEEauthorblockA{
% \textit{Department of Computer Science and Technology} \\
\textit{Tsinghua University}\\
Beijing, China \\
zhangyq24@mails.tsinghua.edu.cn}
\and
\IEEEauthorblockN{Ninghan Zheng}
\IEEEauthorblockA{
% \textit{Department of Computer Science and Technology} \\
\textit{Tsinghua University}\\
Beijing, China \\
zhengnh@tsinghua.edu.cn}
\and
\IEEEauthorblockN{Shanshan Li}
\IEEEauthorblockA{
% \textit{Department of Computer Science and Technology} \\
\textit{Tsinghua University}\\
Beijing, China \\
lishanshan@tsinghua.edu.cn}
\and
\IEEEauthorblockN{Weidong Liu}
\IEEEauthorblockA{
% \textit{Department of Computer Science and Technology} \\
\textit{Tsinghua University}\\
Beijing, China \\
liuwd@tsinghua.edu.cn}
}

\else

\author{\IEEEauthorblockN{1\textsuperscript{st} Given Name Surname}
\IEEEauthorblockA{\textit{dept. name of organization (of Aff.)} \\
\textit{name of organization (of Aff.)}\\
City, Country \\
email address or ORCID}
\and
\IEEEauthorblockN{2\textsuperscript{nd} Given Name Surname}
\IEEEauthorblockA{\textit{dept. name of organization (of Aff.)} \\
\textit{name of organization (of Aff.)}\\
City, Country \\
email address or ORCID}
\and
\IEEEauthorblockN{3\textsuperscript{rd} Given Name Surname}
\IEEEauthorblockA{\textit{dept. name of organization (of Aff.)} \\
\textit{name of organization (of Aff.)}\\
City, Country \\
email address or ORCID}
\and
\IEEEauthorblockN{4\textsuperscript{th} Given Name Surname}
\IEEEauthorblockA{\textit{dept. name of organization (of Aff.)} \\
\textit{name of organization (of Aff.)}\\
City, Country \\
email address or ORCID}
\and
\IEEEauthorblockN{5\textsuperscript{th} Given Name Surname}
\IEEEauthorblockA{\textit{dept. name of organization (of Aff.)} \\
\textit{name of organization (of Aff.)}\\
City, Country \\
email address or ORCID}
\and
\IEEEauthorblockN{6\textsuperscript{th} Given Name Surname}
\IEEEauthorblockA{\textit{dept. name of organization (of Aff.)} \\
\textit{name of organization (of Aff.)}\\
City, Country \\
email address or ORCID}
}

\fi

\maketitle

\begin{abstract}
This full research paper describes the assessment and presentation of student competencies in algorithm courses, grounded in the CC2020 competency model. With the growing emphasis on bridging the gap between academic training and industry demands, competency-based education, which integrates knowledge, skills, and dispositions, has become pivotal in computer science education. To bridge the gap, we need to develop a comprehensive framework to evaluate competencies (knowledge, skills, and dispositions) in computer science education.

The research aims to analyze learning behavior patterns, design methods for competency assessment in algorithm courses, and evaluate the difficulty of course experiments to inform curriculum design. We collected programming experiment and written assignment data from 169 students, adapting it to the xAPI specification for unified analysis. In this work, Markov process modeling was employed to analyze behavioral sequences, revealing cognitive patterns during programming tasks. Multiple methods were applied to quantify competencies (knowledge, skills, dispositions) and identify distinct student clusters. Course difficulty was quantified using proactiveness metrics derived from submission timeliness.

This work contributes a scalable framework for competency assessment in algorithm courses and offers actionable insights for personalized teaching and curriculum optimization. Practically, it enables instructors to tailor interventions based on student clusters and optimize task difficulty. Future work will integrate more students' performance to validate competency models and extend the framework to broader computer science curricula.
\end{abstract}

\begin{IEEEkeywords}
competence, student assessment, multi-modal approaches
\end{IEEEkeywords}

\section{Introduction}
When discussing the way forward for modern computing education, the continuous tracking and research of ACM/IEEE have always been an important position that cannot be ignored. These two authorities have precise insights into the talent gap and needs of the computing industry, and they jointly launched the ``Computing Curricula Guidelines", widely known as Computing Curricula 2005 (CC2005)\cite{10.1145/1121341.1121482}, in 2005, which is of epoch-making significance in that it provides detailed knowledge and learning focuses for each field of computing discipline. This program was groundbreaking in that it provided detailed knowledge domains and learning priorities for each of the computing disciplines, and set a clear direction for computing education around the world \cite{10.1145/3148549}. The report clearly advocates a knowledge-based approach to computing education, which encourages education to be built around a body of knowledge that students already have, to enrich and extend that body as they learn. However, with the passage of time and the gradual advancement of the concept of computer education, this model of education, which favors theory and knowledge, has also revealed some problems.

\begin{figure}[htbp]
\centerline{\includegraphics[width=0.8\linewidth]{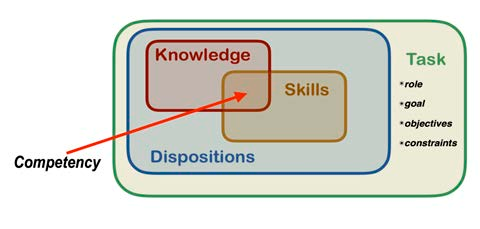}}
\caption{CC2020 Conceptual Structure of the Competency Model\cite{10.1145/3467967}}
\label{fig1}
\end{figure}

In response to the shortcomings of CC2005, ACM/IEEE launched a new reference program for computing education in 2020, Computing Curricula 2020 (CC2020)\cite{10.1145/3467967}, which introduces a compelling educational concept: competency. Compared to the previous version, CC2020 focuses more on industry and society, and emphasizes that the content of the curriculum should be more relevant to the students' future in society \cite{JYJS202304002}. In order to better define and implement this concept, CC2020 introduces a competency model, which not only covers traditional knowledge and skills, but also innovatively adds the dimension of dispositions, forming a complete competency framework of ``Knowledge+Skills+Dispositions" (as shown in Figure \ref{fig1}):

\begin{itemize}
\item \textbf{Knowledge (``know-what")}: computational knowledge, fundamentals and specialized knowledge
\item \textbf{Skills (``know-how")}: the ability to use knowledge to accomplish a task
\item \textbf{Dispositions (``know-why")}: qualities of dispositions in performing tasks
\end{itemize}

The competency-based computing education model is a comprehensive and balanced development model that emphasizes the simultaneous enhancement of knowledge, skills and dispositions. The goal of this education model is clear: to narrow the gap between school education and industry needs, so that students can better integrate and adapt to the ever-changing technological environment and social needs.CC2020 undoubtedly pushes computer education towards a new era focusing on cultivating high-quality computer professionals who are capable of efficiently solving real-world problems. Currently, scholars have conducted in-depth studies and reviews on various aspects of the application of the CC2020 competency model, such as the assessment of skills and dispositions\cite{10.1145/3585059.3611426, 10.1145/3537674.3554747}, visualization of student competencies\cite{10.1145/3328778.3366998, 10.1145/3478432.3499029, gyc}, and how to develop dispositions in computing education\cite{10.1145/3502717.3532121, JYJS202105015, JYJS202207043}.

When examining methodologies for learning data collection, we find that the research projects mentioned above follow their unique methodology in the aspect of data collection and transmission, which ensures that the data can be processed accurately and efficiently in their particular system or study, but also exposes the limitation of the lack of unified normative standards: it is difficult for the data to be seamlessly integrated or sharing. For example, the literature \cite{2021The} introduces an online textbook platform named zyBooks\cite{33/}, which focuses on creating computer science and engineering (CSE) books for network-native audiences. Studies\cite{10.1145/3545945.3569764, 10.1145/3545945.3569732, 10.1145/3626252.3630951, 10.1145/3545945.3569731, 10.1145/3626252.3630847} in computer education have been conducted based on the zyBooks platform, and There are many more platforms like zyBooks \cite{2021The, DHJY202005012, ZDJY201901019, 3/, inproceedings}. In order to facilitate the unified processing and analysis of learning data, we need a standardized framework.

eXperience API(xAPI), once known as Tin Can API, is a  simple, lightweight learning technical specification released in 2013 by the ADL (Advanced Distributed Learning) organization in the US \cite{XJJS201501017}. The main goal is to solve the problem of interoperability of data among different learning systems, which can provide a unified data collection and transmission method, thus enhancing data portability and generalizability. With this standard, learners and educators can gain a more comprehensive understanding of the learning process for more effective learning assessment and improvement \cite{XDYC201405003, ZDJY201901019}. In order to synthesize data from multiple platforms, this study uses the xAPI specification as the basis for data processing and analysis of student learning information. xAPI standard has the following core features:

\begin{itemize}

\item \textbf{Subject-predicate-object structure}: xAPI adopts a subject-predicate-object triple structure to describe learning behaviors, which clearly and concisely records learning behaviors through three elements: actor, verb and object. behavior, as shown in Figure \ref{fig2}.

\begin{figure}[htbp]
\centerline{\includegraphics[width=1\linewidth]{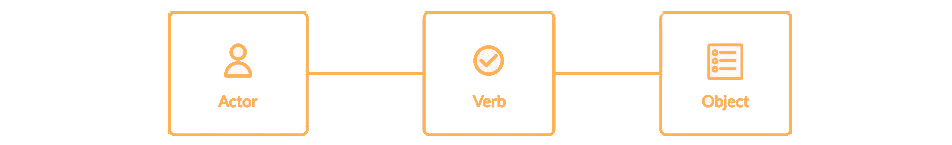}}
\caption{Diagram of the xAPI Phrase Structure}
\label{fig2}
\end{figure}

\item \textbf{Optional Elements}: In addition to the base structure, xAPI includes optional elements such as context and results that provide additional information and documentation of results for learning activities.

\item \textbf{Learning Record Storage System (LRS)}: xAPI defines the LRS to store and manage learning behavior records in a unified format, thus enabling unified archiving, management and analysis of learning data from different systems.

\end{itemize}

\section{Our Algorithm Course}

This study is based on the algorithmic course ``Discrete Mathematics (2)" which focuses on the basic concepts and principles of graph theory and algebraic structures, including the basic concepts of graphs, loop and tree properties and applications, planar graphs and graph coloring, matching and network flows, and the fundamentals of group theory. The course emphasizes the cultivation of learning attitudes and abilities, and provides a variety of performance evaluation modes, including exams, usual assignments, communication of exercise classes and application technical reports, etc. Among them, written assignments and programming lab assignments are usually an important part of the grades, and these two forms of coursework has their own emphasis on the examination of students' knowledge and abilities.

Written assignments deal primarily with the understanding and application of fundamental concepts, theorems, and methods in the course and require students to have a deep understanding and mastery of the core knowledge of discrete mathematics. Assignment problems may cover a wide range of topics from basic conceptual explanations and theorem proving to complex graphical problems and applications of algebraic structures in order to test students' learning in a comprehensive manner.

Programming experiments encourage students to apply what they have learned to solve real-world problems independently, and develop and test their logical thinking. Programming experiments usually require students to understand and implement a certain algorithm in graph theory to deepen their understanding and mastery of the algorithm. Programming experiments focus on examining how students apply the theories and algorithms in discrete mathematics to practical problem solving, such as graph traversal, network flow and other problems. Through programming implementation, students can analyze and verify the data structures in discrete mathematics so as to deeply understand their properties and applications.

The online assessment system for programming experiments can receive programming experiments submitted by students, and automatically assess and give grades according to the test cases. Through the online assessment system, students can get instant feedback on their assignments, and teachers can also check the students' assignment submission status through the system, so that they can know the students' detailed submission history and the progress of their assignments in time. Students can choose between two types of assessments: the ``50\% tests" (covering 50\% of the test cases) and the ``90\% tests" (covering 90\% of the test cases), with a limited number of ``90\% tests" to be taken before the lab deadline. The number of times the ``90\% test" can be taken before the lab deadline is limited, and each time the ``90\% test" is taken, the score of the question is reduced by one point, so students are generally cautious about using the ``90\% test".

In order to ensure that the data can be centrally processed and analyzed, we apply the xAPI specification to the data from different sources in this course. The data collection process first involves the identification of key behaviors, followed by the selection or customization of the appropriate xAPI verbs from the registry and the selection of relevant extensions and resultant data. At the end of each experiment, the administrator uses a crawler to read the behavioral records from the database that have not been uploaded to the LRS, converts them to an xAPI statement and uploads them to the LRS. At this point, the operator (Actor) in the xAPI statement needs to be converted to the student's student number based on the information registered in the experimental platform, so as to make sure that the representation of the operator is uniform across platforms.

Programming experiment data from 169 students was collected from the experimental platform, totaling 10,427 xAPI statements after adapting the xAPI standard. Written assignments data from 174 students were collected and there were a total of 1037 xAPI statements after adapting the xAPI standard.

\section{Behavioral Sequence-Based Analysis Approach}

A Markov Process (MP) is a stochastic process that describes the process of state transfer at discrete time points for a system with a number of states \cite{bhat2002elements}. In this process, the state of the system changes at discrete points in time, and the state transfer at each point in time is related only to the state of the system at that point in time, and not to the state of the system at an earlier point in time. This property is known as ``Markov property" or ``memorylessness". Markov processes depend on the state space of the system and the transition probability matrix (or transition matrix for short). The state space $S=\{S_1, S_2, ..., S_n\}$ is the set of all possible states of the system. The transition probability matrix $P$ describes the probability of transition from any state to any state, specifically, if the system is in state $S_i$ in time $t$, then the probability of the system transition to state $S_j$ in time $t + 1$ is the corresponding element of the transition probability matrix $P_{ij}$ \cite{grigorios2014stochastic}.

This study adopts a simplified, but no less valid, perspective in examining students' behavioral sequences, i.e., it is argued that the primary proactiveness for the decisions students make at a given moment stems to a large extent from the events that occurred immediately before it, i.e., students' behavioral sequences are consistent with Markov property. With the current scale of data available, this simplified analytical approach helps us to understand more intuitively the continuity and dynamics of students' learning behaviors in order to better analyze and predict students' behavioral patterns.

Using the data collected and organized from the experimental platform, this study classifies the states that students are in during the experiment based on their cognitive processes, so that the states include two factors: the students' choices (50\% test or 90\% test) and the results of the experiment (passing all test points, successful compilation but failing to pass all test points, or compilation errors).

As shown in Table \ref{mkv}, in the experiment of Discrete Mathematics (2) course, this study divided the sequence of students' behaviors into five states.

\begin{table}[htbp]
  \centering
  \caption{State Definitions and Frequency}
  \begin{tabular}{l|l|l}
    \hline
    Status & Status Name & Number of state occurrences\\
    \hline
    $S_1$ & 50\% Pass    & 2074 \\
    $S_2$ & 50\% Fail    & 5804 \\
    $S_3$ & 90\% Pass    & 1377 \\
    $S_4$ & 90\% Fail    & 597 \\
    $S_5$ & Compilation Failure & 409\\
    \hline
  \end{tabular}
  \label{mkv}
\end{table}

Our first task is to test whether students' behavioral sequences can actually exhibit Markov property as described above. Define the null hypothesis $H_0$ : The state sequence is not Markovian, according to the Markov property test based on the $\chi^2$ distribution, the $\chi^2$ statistic of the student behavior sequence can be calculated \cite{mkv}.

\begin{figure}[htbp]
\centerline{\includegraphics[width=0.9\linewidth]{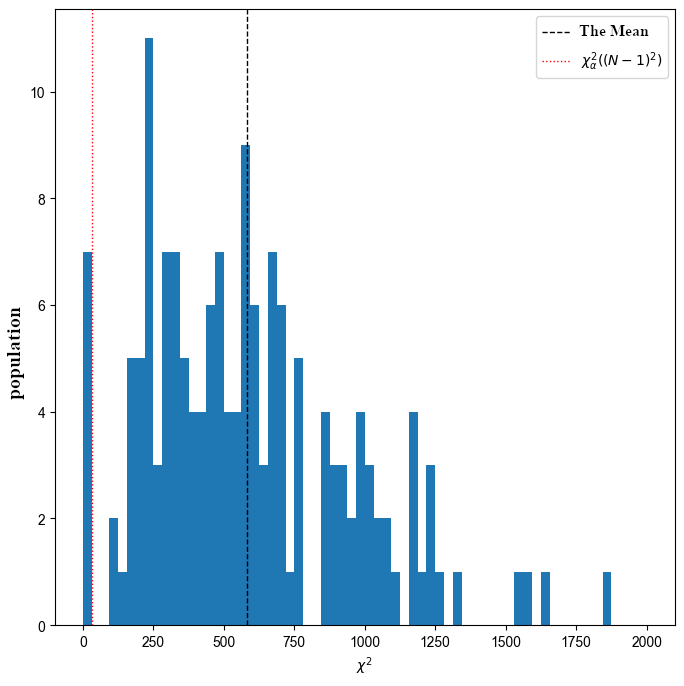}}
\caption{Student Behavior Sequential Markov Property $\chi^2$ Test Results}
\label{fig3}
\end{figure}

The number of states of this model is $N = |S| = 5$. The level of significance is taken as $\alpha = 0.01$ and the critical value is $\chi_\alpha^2((N - 1)^2) = 32.0$. if $\chi^2 > \chi_\alpha^2((N - 1)^2)$, the p-value will be less than the level of significance $\alpha$ and the null hypothesis $H_0$ should be rejected. Calculate the $\chi^2$ statistic of each student behavioral sequence, as shown in Figure \ref{fig3}. It can be seen that the vast majority (more than 95.5\%) of the student behavioral sequences $\chi^2$ indicators exceeded $\chi_\alpha^2((N - 1)^2)$,so the null hypothesis $H_0$ is rejected, and it can be assumed that the student behavioral sequences have Markov property.

Based on the state definitions, there are a total of 10104 state transfers, where the total number of state transfers for each state is shown in Table \ref{mkv2}.

\begin{table}[htbp]
  \centering
  \caption{Frequency of Student Behavioral State Transfer}
  \begin{tabular}{l|p{0.5cm}p{0.5cm}p{0.5cm}p{0.5cm}p{1.2cm}}
    \hline
    \diagbox{Current}{Next} & 50\% Pass & 50\% Fail & 90\% Pass & 90\% Fail & Compilation Failure \\
    \hline
	50\% Pass & 461 & 269 & 1034 & 273 & 35 \\
	50\% Fail & 948 & 4442 & 131 & 118 & 149 \\
	90\% Pass & 390 & 634 & 148 & 28 & 58 \\
	90\% Fail & 169 & 162 & 50 & 166 & 31 \\
	Compilation Failure & 63 & 209 & 9 & 12 & 115 \\
    \hline
  \end{tabular}
  \label{mkv2}
\end{table}

As a whole, students' state shifts had distinct and interpretable statistical features that reflected the overall learning process and the experimental topics. The highest number of transfers from the 50\% Test Failed (50\% Fail) state to their own state (4442, more than four times the frequency of any other state transfer) suggests that once students are in this state, they are likely to remain in it for multiple attempts, demonstrating the process of repeated attempts by students when they have not yet completed their programming assignments. The fact that it was easier to move from the 50\% Pass (50\% Pass) state to the 90\% Pass (90\% Pass) state (1,034 attempts) than to any other state suggests that once a student succeeds in the 50\% Pass test, their program is likely to have completed all of the requirements of the experiment, which is a frequent phenomenon of algorithmic experiments, and for students this feedback during the completion of a topic can help to enhance students' self-confidence in their own code ability, and this phenomenon itself also reflects the logical rigor and low error tolerance of algorithmic topics. It can also be noted that the Compilation Failure state, although not frequent, seems to be more difficult for students to get out of once they enter this state. The relatively low number of transfers from the Compilation Failure state to other successful states may mean that changing one's situation is a relatively large obstacle for students who have made a compilation error, and given that the Discrete Mathematics (2) course is primarily geared towards first-year students who have not long been in the university, some students with a weak foundation in programming sometimes repeatedly fall into Compilation Failure situations, and may need additional help from the teaching team.

Due to the significant Markov property of student behavioral sequences, a Markov process-based behavioral model can be constructed for each student's learning path in the programming experiment. In this model, the probability of a student transition from one learning state to another is the core of the model construction.

\begin{figure}[htbp]
  \centering
  \subcaptionbox{Student A\label{fig:mkv6-a}}
    {\includegraphics[width=0.48\linewidth]{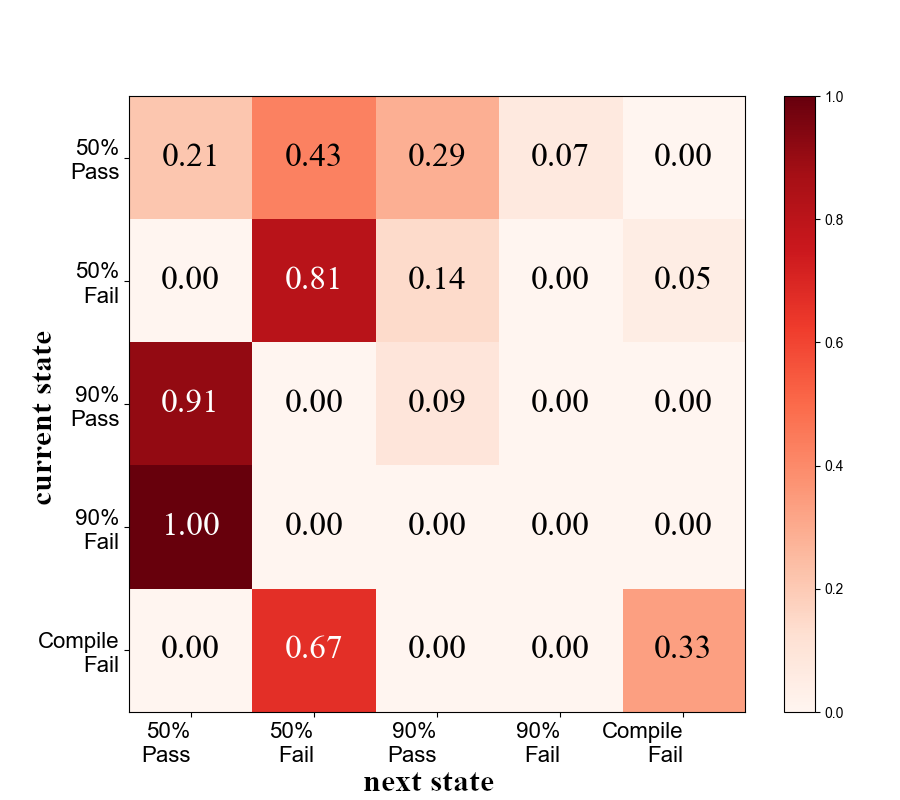}}
  \subcaptionbox{Student B\label{fig:mkv6-b}}
    {\includegraphics[width=0.48\linewidth]{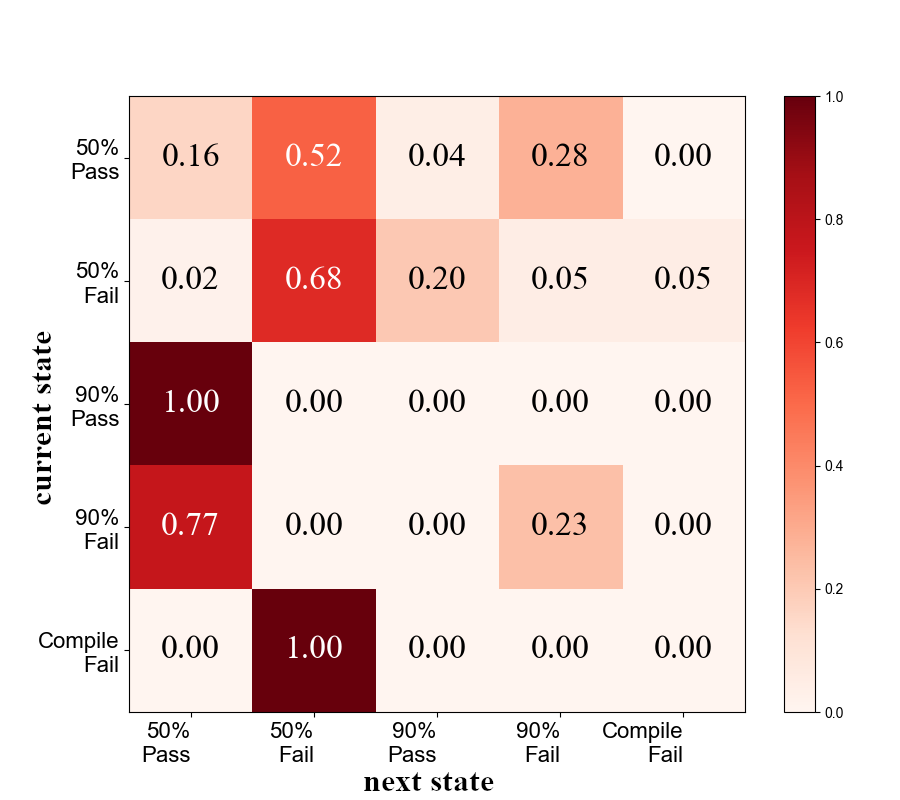}}
  \caption{State Transition Matrices for Two Typical Student Behavior Sequence Models}
  \label{fig4}
 \end{figure}

Markov-based behavioral models can be fully described by this state transition matrix, and an in-depth analysis of each student's state transition probability distribution can not only reveal the unique behavioral patterns of students in the learning process, but also provide insight into their cognitive dispositions and preferences. The behavioral sequence models of two typical students are visualized and their state transition matrices are shown in Figure \ref{fig4}.

Student A has a higher probability of continuing to have failed 50\% tests after failing the 50\% test. The tendency to rewrite and resubmit unfinished code when the code fails to achieve the expected results may be an indication of Student A's exploratory strategy in problem solving, as Student A may be trying to guess the criteria of the assessment system in order to adjust his code to meet the requirements of the assessment. While this behavior may sometimes help students find ``shortcuts" to passing the assessment, it may also cause students to lose sight of the importance of understanding and grasping the details of the algorithm.

The probability of Student B falling into the code compilation failure state is relatively low, and he is also able to get out of the compilation failure state immediately, reflecting that Student B has a relatively skillful mastery of the programming language, and he is able to avoid common compilation errors more accurately in the writing process. However, the probability of Student B moving from the 50\% test passed state to the 90\% test failed state is higher, and his code has problems in further testing, which may indicate that Student B's meticulousness in writing the code needs to be improved or that he does not have a full understanding of the algorithms taught in the course.

\section{multidimensional data-based methods to analyzing student competencies}

\subsection{Principal Component Analysis}

According to the CC2020 competency model departure, student competency has multiple dimensions of performance, so we need to integrate data from multiple sources in order to conduct a more comprehensive and integrated analysis of student competency. Since the original dataset contains a variety of competency-related and non-competency-related features in the students' learning process, this study extracts the main and representative indicators from the original data through the data dimensionality reduction technique in order to more accurately and quantitatively assess a certain competency feature of the students.

Principal Component Analysis (PCA) is a dimensionality reduction technique widely used in data analysis and statistics. PCA transforms the original high-dimensional data into a set of linearly independent representations of each dimension by linear transformation to identify and extract the main feature components in the data, so as to achieve dimensionality reduction of the high-dimensional data and highlight the key information in the data. Selecting some large principal components to be retained can effectively condense the information in the data and help to understand and analyze the complex data set more clearly \cite{mackiewicz1993principal}.

We apply the PCA method to downscale the data reflecting this disposition to one dimension, in order to serve as a quantitative indicator of this disposition of students. For the purpose of student dispositions analysis, this study first selected three of the 11 dispositions from the list of dispositions given by CC2020 that were more closely related to the study data and defined the trait dimensions that had a more pronounced effect on each trait, as shown in Table \ref{feature}.

\begin{table}[htbp]
  \centering
  \caption{Competency Dispositions Corresponding to Data Sources}
   \begin{tabular}{p{1.5cm}|p{2.5cm}|p{3cm}}
       \hline
       Hallmark & CC2020 Definitions & PCA Source of data\\
       \hline
       Knowledge  & know-what & Overall performance\\
       \hline
       Skills & know-how & Experimental results\\
       \hline
       Meticulous & Attentive to detail; accurate, thoroughness & Overall pass rate for lab reviews\\
       \hline
       Proactive & With initiative, self-starter, independent & Lab completion time, written assignment submission time\\
       \hline
       Passionate & Conviction, strong commitment, compelling & Total number of experimental submissions, behavioral model counts for each type of behavior\\
       \hline
   \end{tabular}
   \label{feature}
\end{table}

Using the above methodology, this study conducted a competency analysis of students taking Discrete Mathematics (2) and obtained the results of the two basic dimensions of knowledge and skills as well as three dispositions. The analysis results for each dimension were mapped to a real number between 0 and 1, with larger values indicating better performance on that dispositionsization dimension.

\begin{figure}[htbp]
  \centering
  \includegraphics[width=1\linewidth]{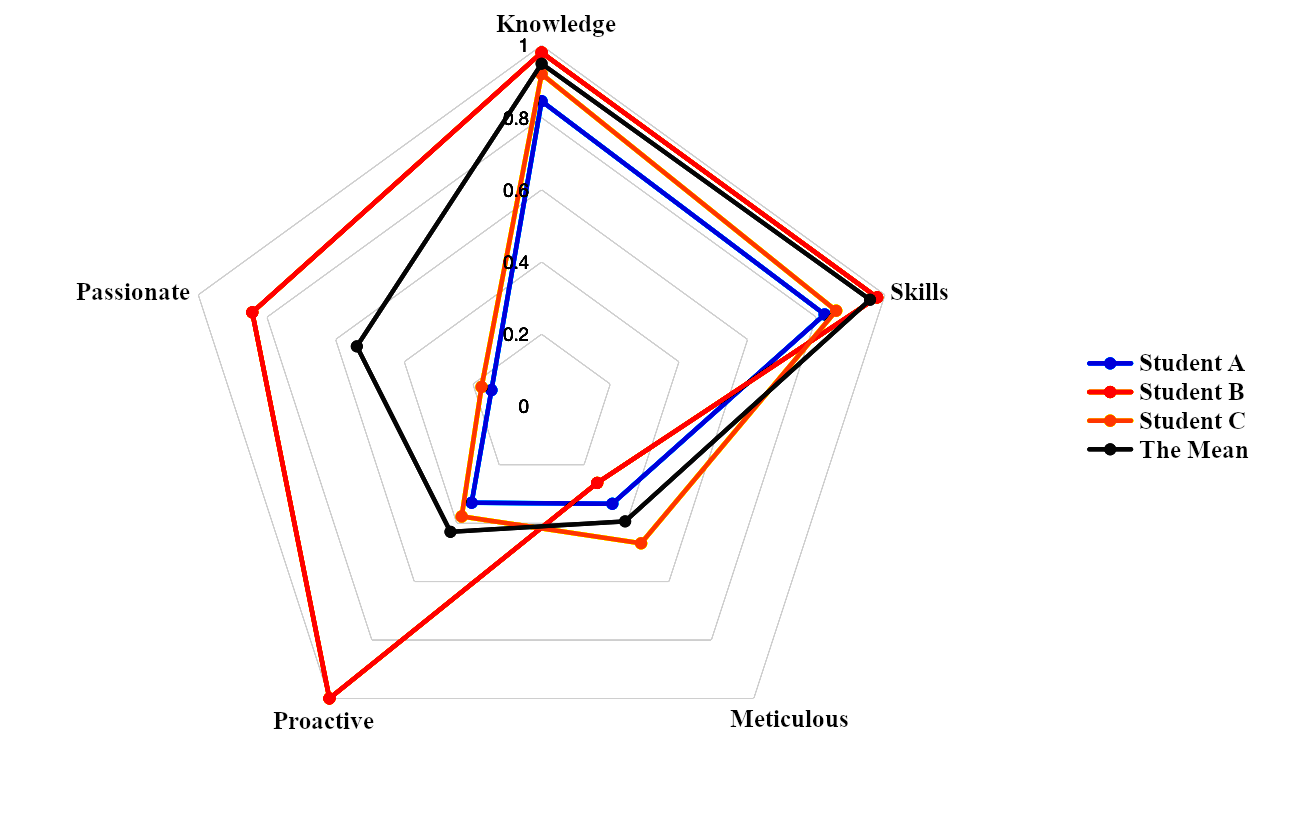}
  \caption{Student Competency Analysis Visualization Results}
  \label{spider}
\end{figure}

In order to show the specific situation of students' competence more clearly, the results of students' competence analysis can be shown in the form of a spider web diagram, in which the data of three representative students are selected as an example, as shown in Figure \ref{spider}.

The student competency profile in Figure \ref{spider} can be analyzed as follows.

\textbf{Student A}

Student A's proactiveness, enthusiasm, and poor performance on the knowledge and skill dimensions reflect the student's possible lack of proactiveness and interest in the algorithms class and, ultimately, his or her failure to achieve a relatively good grade. The student generally started experiments late, primarily driven by deadlines, and his or her scheduling of assignments and experiments was not reasonable enough, and he or she was relatively lacking in initiative. The data in the other dimensions show that the student is more detail-oriented in the experiment process. In the teaching process, it can be prompted to arrange the time of homework in advance, use more sufficient time to better complete the homework, and strengthen the accumulation of knowledge in the daily course learning, or take the initiative to provide some help to them.

\textbf{Student B}

Student B performed relatively well on both the Knowledge and Skills traits and relatively poorly on the Attention to Detail Dispositions trait. This indicates that the student finally achieved relatively good grades in both written assignments and experiments, but the phenomenon of submitting incorrect code and then fixing it several times during the experimental process made the overall passing rate of the experimental assessment low, which shows that the student may lack attention to details in completing the experiments, preferring to write problematic code and then fixing it according to the feedback from the online assessment platform. In the teaching process, according to this disposition, we can prompt him to pay attention to the details of the experiment, and strive to avoid errors in the design and implementation of the algorithm.

\subsection{Student Competency Clustering Analysis}

Further observation of students' competency performance reveals that the competency performance of many different students has a fairly high degree of similarity, and students with similar dispositions can be categorized into a group to more accurately grasp the learning dispositions of different groups of students and to provide teaching suggestions with a more holistic perspective in order to promote the all-round development of students \cite{10.1145/3545945.3569882}.

In this study, clustering is performed using the K-means algorithm, a classical clustering algorithm that aims to divide n observations into k subgroups (or clusters) such that each of them belongs to the cluster corresponding to its nearest mean (i.e., the center of the cluster). The K-means algorithm is a simple and efficient way to make the data points within the same cluster as similar as possible, and data points between different clusters as different as possible \cite{macqueen1967some}.

\begin{figure}[htbp]
  \centering
  \includegraphics[width=0.875\linewidth]{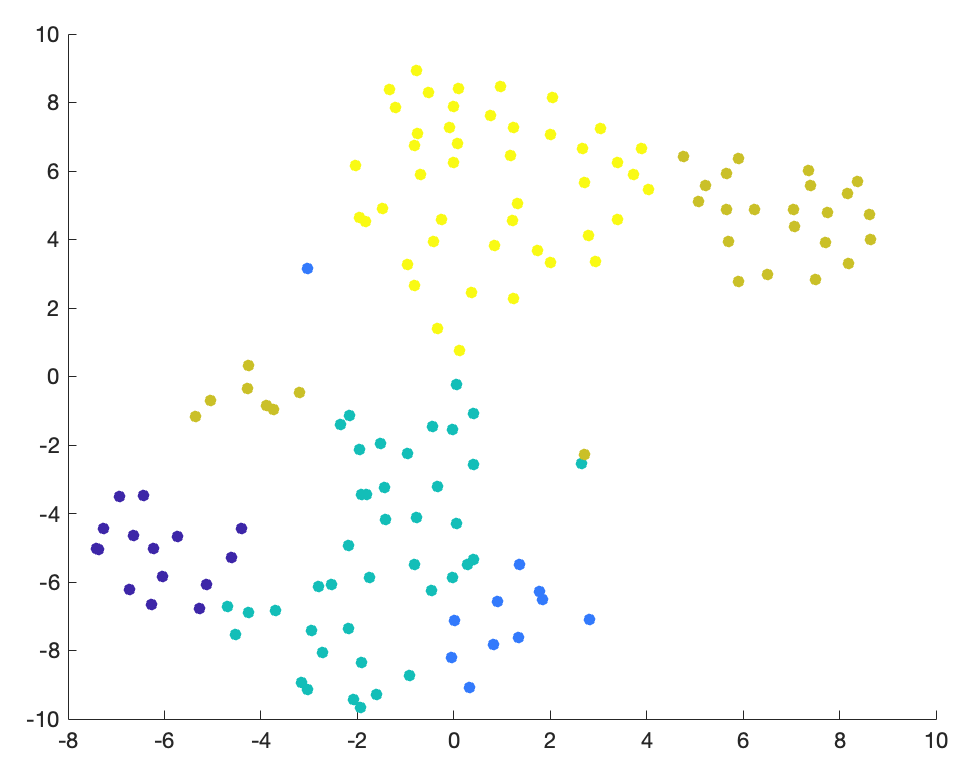}
  \caption{Visualization of the Effect of Clustering Algorithm Based on t-SNE Algorithm}
  \label{tsne}
\end{figure}

We can verify the clustering effect of K-means using visualization. t-Distributed Stochastic Neighbor Embedding (t-SNE), which is an efficient method that can map high-dimensional data to a low-dimensional space in a nonlinear way, can try to preserve the similarity relationship between data points \cite{2008Visualizing}. Figure \ref{tsne} shows the data and clustering results after downscaling the student competency features to two dimensions by using t-SNE, where different colors indicate different groupings, which visually demonstrates the inner structure of the data and initially verifies the effectiveness of K-means clustering.

\begin{figure}[htbp]
  \centering
  \includegraphics[width=1\linewidth]{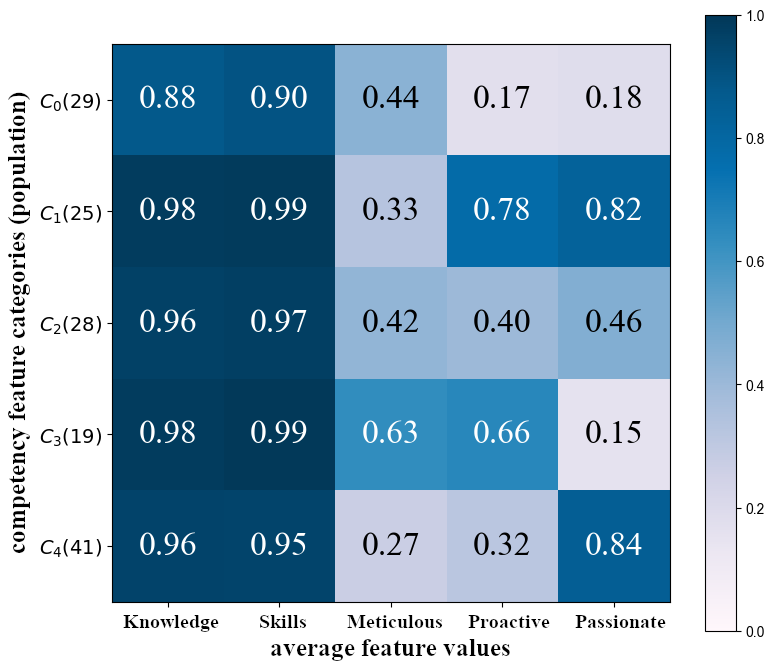}
  \caption{Clustering Results Data}
  \label{clu}
\end{figure}

Figure \ref{clu} shows the resultant data of the K-means algorithm clustering, where each row represents a subgroup (cluster), each row represents a competency feature, and the numbers and colors in the cells show the average feature values for that subgroup.

By looking at the Figure \ref{clu}, we can see that there is significant variability between the different categories of students, and based on the average disposition values, we can generalize the more distinctive behavioral dispositions of each category.

\begin{itemize}
\item Subgroups $C_0$ of about 20\% of the students are usually slightly less advanced in knowledge and skills, and both proactiveness and enthusiasm are relatively low. Students in this category may appear to be lagging behind in learning knowledge and skill acquisition due to a lack of some proactiveness or lack of interest in programming experiments. These students may need more proactiveness or help in solving difficult obstacles in their learning process.
\item The students in the subgroups $C_1$ have achieved a high level of knowledge and skills. They are highly motivated, although they may not be fine-tuned in their attention to detail. These are likely to be the students who, although they may feel some challenges in their learning, are able to complete tasks first time with a high level of enthusiasm for learning.
\item The students in the subgroups $C_2$ were well balanced in all areas of performance, and although they did not show a very high level of proactiveness or enthusiasm in the course, they were able to rationalize the time taken to complete their assignments and achieve better grades.
\item The number of students in this subgroup $C_3$ was significantly smaller than in the other subgroups, and this group of students demonstrated excellent knowledge and skills in the coursework and experiments, not only being motivated to learn, but also being able to pay sufficient attention to detail to be able to handle the various topics in the programming experiments with ease. It is reasonable to assume that students in this subgroup have strong academic skills or have studied the content of this course systematically before. This group of students was slightly less enthusiastic, most likely because the topics in the programming lab were not challenging enough for them.
\item The subgroup $C_4$ has the largest number of students, and such students still have room for improvement in their knowledge and skills, and do not tend to finish their assignments at the first time after they are assigned, but they still have a high level of enthusiasm for programming and algorithm learning. Their ability to pay attention to details in programming experiments needs to be improved, and they may make repeated trial and error and constant revisions in the experiments, but they can eventually master the content taught in the course.
\end{itemize}

By constructing a competency model and conducting a cluster analysis, we can gain a deeper understanding of students' individual differences and learning needs. It is worth noting that the dispositions of students' learning are relatively stable over a period of time, and once we have identified students' cognitive styles, we can predict to a certain extent how they will perform in their future learning. This not only helps to realize personalized teaching, but also provides a scientific basis for teachers' teaching strategies, and ultimately achieves an overall improvement in teaching quality.

\section{Competency Analysis and Curriculum Design}

\subsection{Analysis of the Correlation Between Proactive Behavior and Level of Competence}

The topics of each programming experiment may be divided into different groups, and only 1 topic belonging to the same group will be allowed to be selected; more than 1 part will not be marked as the final version and thus will not be scored. In the Discrete Mathematics (2) course experiment, each group of questions examines the same algorithm, with a small difference in difficulty, but progressively more difficult in numbered order, with a corresponding increase in the value of the questions. Proactive dispositions in competency can provide a contrast between experiments in different groups, because proactive dispositions reflects the students' time schedule for completing the experiment; the more difficult the topic, the longer it will take for students to solve it, and the more interruptions from other matters will increase in the process, which will ultimately show up as a general decrease in students' proactive dispositions on the topic. To explore the relationship between competency and curriculum design, we further analyzed students’ proactive behaviors as a key indicator. The reason this study did not use student grades directly to reflect the difficulty of the experiment is that the vast majority of students in the programming experiment received near perfect scores with very little variance, which makes the grade data less informative than it should be, whereas the distribution of proactive dispositions behaviors is relatively even and better represented.

\begin{figure}[htbp]
  \centering
  \includegraphics[width=1\linewidth]{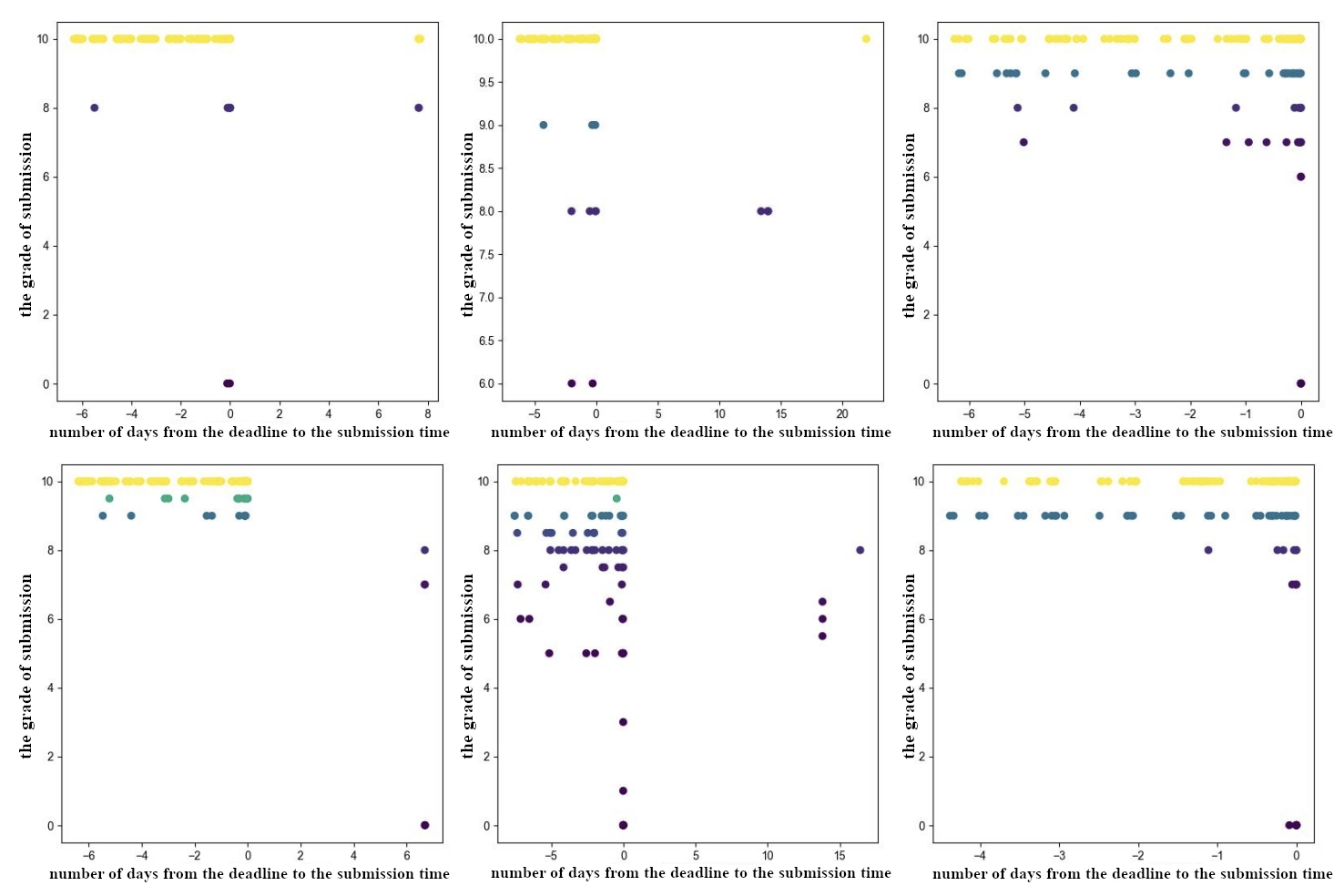}
  \caption{Graph Analyzing the Relationship Between Time of Submission of Written Assignments and Grades}
  \label{allhw}
\end{figure}

First of all, the correlation between proactive dispositions and proficiency level was analyzed in this study to confirm that proactive dispositions does reflect students' learning. According to the table \ref{feature}, proactive dispositions depends on the time of written assignment submission, making a graph of the relationship analysis between the time of submission of each written assignment and the grade, as shown in the Figure \ref{allhw}.

It is clear from the graph that students who tend to be very proactive in completing assignments, and who complete their assignments long before the written assignment deadline, tend to do relatively well; students who do relatively poorly in their courses are more likely to rush to meet assignment deadlines or even turn in their assignments late, suggesting that there is a strong correlation between being proactive in written assignments and a student's level of competence.

Next, we quantitatively define and analyze a classmate's Proactiveness in completing a task. First, by analyzing the student's assignment completion time, we can quantify a classmate's activeness in completing an assignment as the ratio of the time difference between the assignment completion time and the assignment deadline time to the time difference between the start time and the deadline time of that assignment, yielding

\begin{equation}
\text{Proactiveness}_{ij} = \frac{\text{Deadline}_i - \text{CompletionTime}_{ij}}{\text{Deadline}_i - \text{StartTime}_{i}}
\end{equation}
\begin{itemize}
    \item $\text{Deadline}_i$ indicates the deadline for the assignment $i$
    \item $\text{CompletionTime}_{ij}$ indicates when the student $j$ completed the assignment $i$
    \item $\text{StartTime}_i$ indicates the start time of the assignment $i$
    \item $\text{Proactiveness}_{ij}$ quantifies the proactiveness of student $j$ in completing assignment $i$
\end{itemize}

\begin{figure}[htbp]
  \centering
  \subcaptionbox{Experiments Proactiveness and Performance \label{aa}}
    {\includegraphics[width=0.49\linewidth]{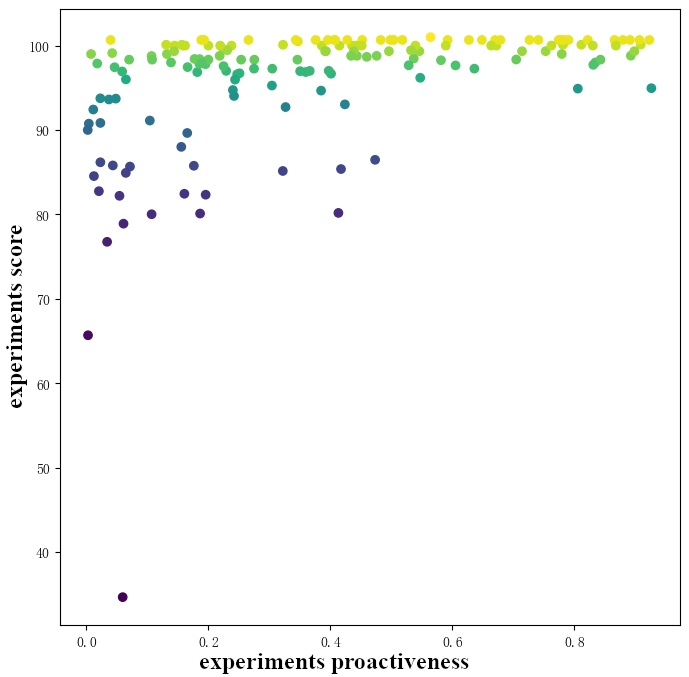}}
  \subcaptionbox{Overall Proactiveness and Performance\label{ab}}
    {\includegraphics[width=0.49\linewidth]{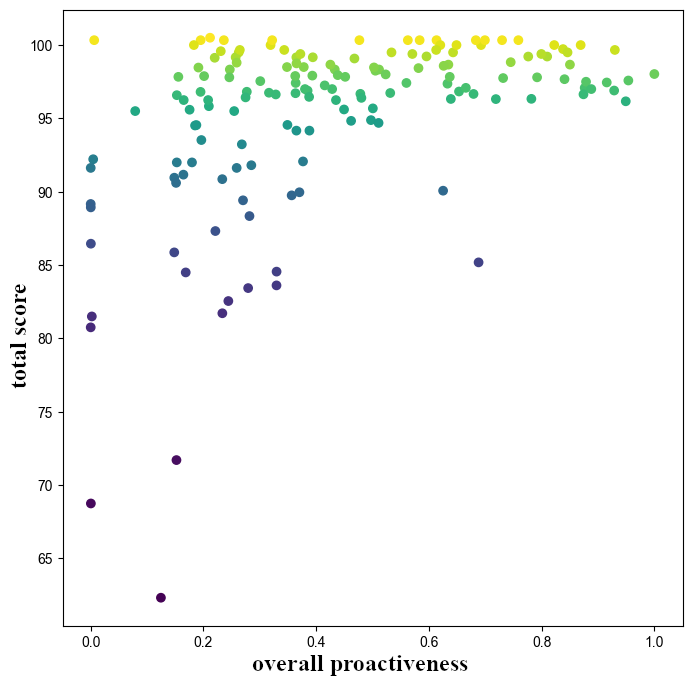}}
  \caption{Correlation Analysis Between Proactiveness and Performance}
  \label{fig9}
 \end{figure}

For programming experiments, we can analyze the relationship between experimental proactiveness and experimental performance, as shown in Figure \ref{aa}; for written and programming assignments overall, we use the average level of proactiveness as an estimate of the overall level of proactiveness, resulting in an analytical scatterplot of the relationship between overall proactiveness and overall performance, as shown in Figure \ref{ab}.

 It is obvious from the graphs that the degree of motivation has a high correlation with the level of competence in both the programming experiments and the assignments in general, which suggests that our choice of the degree of motivation as an evaluation index reflecting the difficulty of the experiments is reasonable. In addition, there is no scatter distribution in the lower right corner of both graphs, which indicates that students with a higher degree of positivity will not achieve poorer grades in their assignments, and helps to motivate motivated students to continue to work hard in this course.
 
\subsection{Quantification of Difficulty Based on the Degree of Experimental Positivity}

We further noted that the proactive dispositions of the students was not only determined by their own factors, but also correlated with the difficulty of the course experiments. By analyzing the proactive behavior of the students, we can invert whether the course experiment design is too easy or too difficult.

We define the difficulty of a set of lab assignment questions as the distance between a distribution consisting of the Proactiveness of all students who completed the set of questions on time and a uniform distribution $U[0, 1]$. Let \(\text{Proactiveness(Q)} = \{a_1, a_2, \ldots, a_n\}\) be the set of Proactiveness of all students who completed the questions Q on time, and these Proactiveness values are sorted in non-decreasing order. Then generate a uniform distribution \(U = \left\{ \frac{k}{n} \right\}_{k=0}^{n-1}\) with the number of elements $N$.

Define the difficulty of the assignment Q, \(D(\text{Q})\), as the standardized Euclidean distance between the distribution of positivity and the uniform distribution, calculated by the formula

\begin{equation}
D(Q) = \frac{1}{\sqrt{n}} \left\| \text{Proactiveness}(Q) - U \right\|_2
\end{equation}

where \(\left\| \cdot \right\|_2\) denotes a two-paradigm number, i.e., a Euclidean paradigm.

\begin{figure}[htbp]
  \centering
  \includegraphics[width=0.8\linewidth]{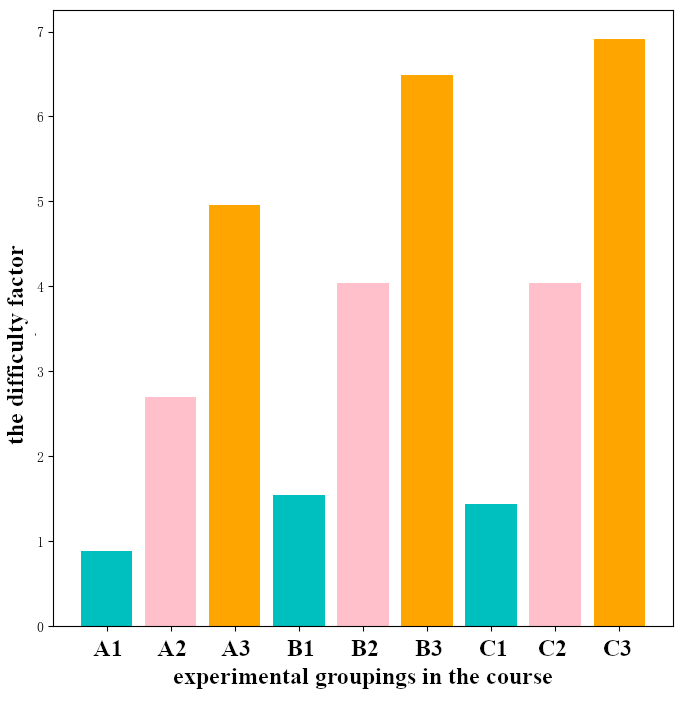}
  \caption{Quantifying the Difficulty of Experiment Topic Sets Based on Proactiveness}
  \label{acthard}
\end{figure}

The results of quantifying the difficulty of the experimental groupings for the Discrete Mathematics (2) course are shown in Figure \ref{acthard}. It can be seen that the final quantitative results obtained for the difficulty factor of the course experiments are as expected: as the course progressed, the difficulty of each programming assignment (three-color loop) increased in comparison to the previous one.

\section{Conclusions and discussion}

The purpose of this paper is to carry out research on methods for assessing and demonstrating student competency based on the algorithmic course Discrete Mathematics (2). Specifically, based on the CC2020 competency model and related work, the main findings of this paper are summarized as follows:

\begin{itemize}
    \item Completed the collection of learning data and xAPI specification adaptation for the course ``Discrete Mathematics (2)", constructed a unified processing and analysis capability for learning data, and laid the foundation for subsequent research.
    \item A Markov process-based model for analyzing students' behavioral sequences was constructed to reveal the behavioral patterns and their underlying cognitive dispositions during the learning process by modeling the behavioral sequences of students' programming experiments.
    \item A student competency analysis based on multidimensional data was carried out, using principal component analysis (PCA) to achieve effective quantification of student competency dispositions, and clustering algorithms to reveal differences in learning styles of student groups, which helps to provide decision support for personalized teaching.
    \item A course design analysis based on student competency analysis is proposed, which provides a reference for improving the teaching quality of the course by quantitatively evaluating the factors of experimental difficulty.
\end{itemize}

It is hoped that students' other grades can be incorporated into the competency assessment model in order to more comprehensively assess students' knowledge of courses and to verify whether our clustering can be more significantly correlated with students' grades \cite{10.1145/3545945.3569764}.

We look forward to more exchanges and collaborations with other programs to explore more comprehensive methods of assessing and demonstrating competence and to make more new advances in the field of computer education.

\bibliographystyle{IEEEtran}
\bibliography{IEEEabrv,refs}

\end{document}